# PHOTOMETRY OF AN UNUSUAL SMALL DISTANT OBJECT 2016 ND21

Hromakina T. A.[1], Velichko S. F.[1] Belskaya I. N.[1], Krugly Yu. N.[1], Sergeev A.V.[2,3]

[1]Institute of Astronomy, V.N. Karazin Kharkiv National University, Sumska str. 35, Kharkiv 61022, Ukraine;
[2]IC AMER, National Academy of Sciences of Ukraine, Akademika Zabolotnoho str. 27, 03680 Kyiv, Ukraine,
[3]Terskol Branch of INASAN, Russian Academy of Sciences, 81, Elbrus ave., ap. 33, Tyrnyauz, Kabardino-Balkaria Republic, 361623, Russian Federation.

Contact author: hromakina@astron.kharkov.ua

**Abstract:** We present the first-ever measurements of the surface colors and rotational properties of a recently discovered distant object in an unusual orbit 2016 ND21. In 2017, this object was close to its perihelion, which allowed us to observe it with a 2.0-m telescope at the Peak Terskol observatory. Broadband photometric observations of 2016 ND21 were carried out in October and December of 2017 in standard BVR filters of the Johnson-Cousins photometric system. During our observations, we did not detect any cometary activity. We found the rotational period of 17.53±0.02 hr, while another a little longer value of 17.65±0.02 hr is also possible. Assuming an equatorial aspect of observations, a peak-to-peak amplitude of A = 0.31±0.05 mag (or even higher since only one maximum and one minimum were well-measured) corresponds to an elongated body with an axis ratio a/b ~1.3. The lightcurve behavior indicates a complex, possibly non-convex, shape of this object. The visible absolute magnitude is $H_V$ =12.4±0.1 mag, which was estimated by using the linear phase slope 0.04 mag/deg as the most probable value from our observations. This phase slope suggests a low-albedo surface of 2016 ND21. Assuming a surface albedo in the range of 0.04-0.10, the size of 2016 ND21 should be about 15-23 km. From our multi-color observations, we determined surface colors V-R = 0.69±0.04 mag, B-R = 1.79±0.08 mag, and B-V = 1.10±0.08 mag. The measured colors indicate an extremely red surface of this object. A very red surface is unusual for comets, which is in agreement with the fact, that no cometary activity was detected for 2016 ND21. The B-R color is higher than the typical B-R colors of the red D-type asteroids, but it is consistent with colors of the red Centaurs and TNOs classified as RR type in TNOs classification. This result gives a first evidence of a possible outer belt origin of this small body.

**Keywords:** Minor planet, photometry, lightcurve, rotational period, surface colors

## Introduction

Recently discovered small Solar system object 2016 ND21 has an unusual high-eccentric ($e$ = 0.55) and highly inclined ($i$ = 21.83 deg) orbit with a major semi-axis $a$ = 8.45 AU. Following the classification described by Gladman et al. (2008), this object with a perihelion $q$ = 3.76 AU and a Tisserand parameter $T_J$=2.584 should be classified as a cometary body. However, the orbit's inclination of 2016 ND21 is too high for a Jupiter Family Comet member (e.g. Lowry et al. 2008), and no cometary activity was discovered for this object so far. Jupiter Family Comets, although being a dynamically distinct group, are similar to some Centaurs, small bodies orbiting between Jupiter and Neptune. The object 2016 ND21 cannot be classified as a Centaur because a perihelion of its orbit is in asteroid belt. In the Minor Planetary Center (MPC) catalog (https://minorplanetcenter.net/iau/lists/Unusual.html), 2016 ND21 is classified as a distant object with an unusual orbit. Its aphelion is Q=12.70 AU. There are only a few discovered objects of this kind, and one of them is 2008 SO218, that is considered to be in a retrograde resonance with Jupiter (Morais et al. 2013).The objects at such unusual orbits are believed to be captured by Jupiter and can originate from both, the inner or the outer parts of our Solar system. Studying the physical properties of these objects can shed a light on their origin. No size or albedo measurements are available for 2016 ND21. In this paper, we present the first-ever measurements of the surface colors and rotational properties of this object.

**Observations and data reduction**

The observations of 2016 ND21 were carried out during six nights in October and December of 2017 at the 2.0-m telescope of the Peak Terskol observatory. The telescope is equipped with a 2048 × 2048 CCD camera FLI PL-4301, that has a pixel scale 0.31arcsec per pixel, resulting in a field of view 10.7 × 10.7 arcmin.
Observations were performed in BVR filters of the standard Johnson-Cousins photometric system. At the moment of observation, the object was near its perihelion and had visible magnitude within the range 18-19 mag. During each observation run, we also acquired bias, dark, and flat-field images. Table 1 shows the observational circumstances, namely heliocentric (r) and geocentric ($\Delta$) distances, Solar phase angle ($\alpha$), mean values of reduced magnitudes in B, V, and R filters, and a total observing time ($T_{obs}$) in hours.

Table 1. Observational circumstances and measured reduced magnitudes in BVR filters for 2016 ND21

| UT time | r, AU | $\Delta$, AU | $\alpha$, deg | B(1, $\alpha$), mag | V(1, $\alpha$), mag | R(1, $\alpha$), mag | $T_{obs}$, hr |
|---|---|---|---|---|---|---|---|
| 2017 10 12.80 | 3.770 | 2.993 | 10.65 | - | - | - | 2.4 |
| 2017 10 13.77 | 3.771 | 3.001 | 10.64 | - | - | - | 6.1 |
| 2017 10 20.66 | 3.775 | 3.065 | 11.79 | - | 12.790±0.03 | 12.023±0.03 | 1.0 |
| 2017 12 10.72 | 3.816 | 3.727 | 14.94 | 14.117±0.08 | 13.022±0.05 | 12.218±0.03 | 3.9 |
| 2017 12 12.76 | 3.818 | 3.756 | 14.91 | 14.258±0.08 | 13.190±0.04 | 12.437±0.04 | 2.9 |
| 2017 12 13.75 | 3.819 | 3.770 | 14.89 | 14.121±0.09 | 13.065±0.05 | 12.223±0.05 | 0.8 |

Data reduction was made using MaximDL software package in the standard way, which included bias and dark-field subtraction from the raw data and flat-field correction. Differential photometry was done using the ASTPHOT package that was developed at DLR by S. Mottola (Mottola et al. 1995).The accuracy of differential photometry was about 0.02-0.03 mag in V and R filters and ~0.07 in B filter. For two first nights in October, only differential photometry was performed. However, the same reference star was used to connect those two lightcurves. On October 20, December 12 and 13, the standard star GD246 from Landolt catalog (Landolt 1992) was observed, which allowed us to perform absolute calibrations for four out of six nights. The uncertainty of absolute photometry was 0.04 mag in V and R filters, and 0.08 mag for B filter. Thus, an absolute calibration and using the same reference stars for the differential photometry allowed us to exclude an arbitrary shift of the data while searching for the rotation period.

**Results and discussion**

We inspected the obtained images of 2016 ND21 on the existence of cometary coma by comparing its profile with those of the field stars. No broadening of the object's PFS that could suggest a cometary activity was detected.

The search for a rotational period was made following the Fourier analysis technique described in Harris& Lupishko (1989) and Magnusson (1990). The observations during two consistent nights on 12 and 13 of October revealed that the period should be longer than ~16 hr. Using all obtained data, we found the rotation period P = 17.53±0.02 hr. We also noted that another value of period 17.65±0.02 hr is also possible. The composite lightcurve with period P=17.53±0.02 hr is shown in Fig. 1. The estimated amplitude is A = 0.31±0.05 mag, but it can be even higher since two extrema of the lightcurve are not covered at their extreme values. The measured amplitude corresponds to an elongated body with an axis ratio a/b ~1.3, if we assume an equatorial aspect of observations and ellipsoidal shape. However, the lightcurve has a rather unusual behavior with a broad minimum and a sharp maximum, which suggest that 2016 ND21 have not a simple ellipsoidal shape, but rather a complex, possibly non-convex, one.

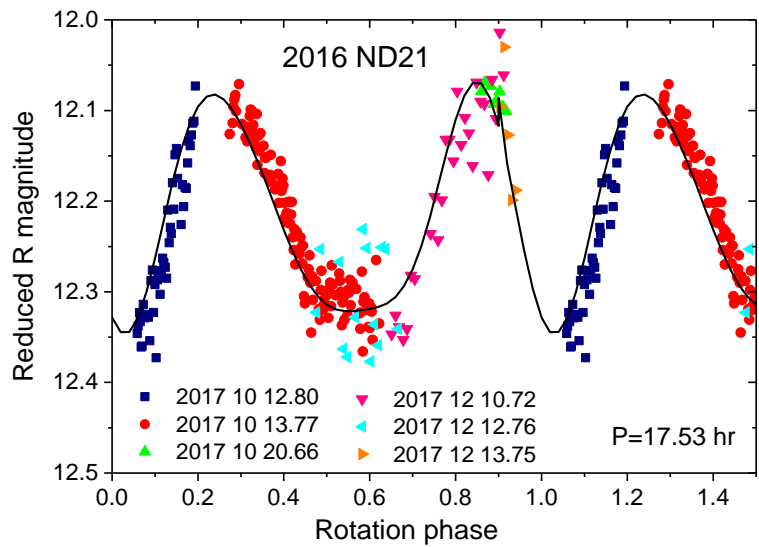

Fig. 1. Composite lightcurve of 2016 ND21 with rotational period 17.53 hr.

From our multi-color observations, we determined surface colors V-R = 0.69±0.04 mag, B-R = 1.79±0.08 mag, and B-V = 1.10±0.08 mag. The obtained colors reveal extremely red surface of this object. Comet nuclei tend to have more neutral colors (average V-R=0.49±0.12 mag compared to average V-R=0.59±0.08 mag for Trans-Neptunian objects (TNOs); Snodgrass et al. 2006), that according to Jewitt (2002) may be due to resurfacing caused by comet activity. In this regard, the absence of cometary activity of 2016 ND21 is in agreement with its very red surface. The B-R color is higher than a typical color of the red D-type asteroids (B-R = 1.22±0.09; Bus & Binzel 2002, Fornasier 2007), but falls within the color range of red Centaurs and TNOs classified as RR type in TNOs' classification (Barucci et al. 2005).We converted the measured colors of 2016 ND21 to spectral reflectance and compared them with the average spectra for D-asteroids and RR Centaurs and TNOs. Fig. 2 presents mean RR-type spectrum based on N=11 objects (Merlin et al. 2017) and mean D-type spectrum acquired using N=9 asteroids (Bus & Binzel 2002) in comparison with our data. A similarity of spectral slope of 2016 ND21 to red Centaurs and TNOs gives a first evidence of a possible outer belt origin of this object.

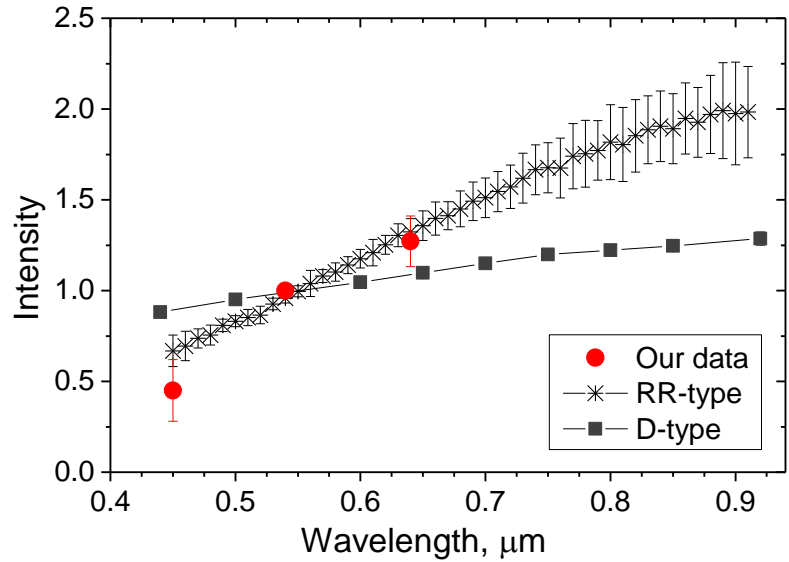

Fig. 2. Comparison of our data with mean RR-type (Merlin et al. 2017) and D-type (Bus & Binzel 2002) spectra.

Our data covered the phase angle range from 11.8 to 14.9 deg. We tried different phase slope coefficients while building the composite lightcurve and found that the most probable value is about 0.04 mag/deg, which is characteristic to a dark surface (see e. g. Belskaya & Shevchenko 2000). Using this phase coefficient and assuming linear phase-angle dependence, we estimated a value of visible absolute magnitude $H_V$ =12.4±0.1 mag. This value is in a good agreement with that from MPC catalog ($H_V$ ~12.3 mag). In assumption of a low albedo surface (0.04-0.1), the diameter of 2016 ND21 will be from ~15 km to ~23 km.

**Conclusions**

We report results of the first photometric investigation of the unusual distant object 2016 ND21. Although this object has a cometary orbit, as per our data it does not reveal any signs of cometary activity. Two values of rotational periods P = 17.53±0.02 hr and P = 17.65±0.02 hr were found, the former being more preferable. The unusual behavior of the lightcurve suggests a complex and possibly non-convex shape of 2016 ND21. Assuming an equatorial aspect of observations, the peak-to-peak lightcurve amplitude A=0.31 mag corresponds to an elongated body with axis ratio a/b ~1.3. The object 2016 ND21 has a very red surface similar to RR type in TNOs' classification, which suggests a possible outer belt origin of this small body. We defined its visible absolute magnitude $H_V$ =12.4±0.1 mag using the linear phase slope 0.04 mag/deg as the most probable value from our observations. The phase slope suggests a low-albedo surface of this object. In assumption of an albedo from 0.04 to 0.1, the size of 2016 ND21 should be within 15-23 km.

Thus, our BVR photometric observations of a small distant object 2016 ND21 at an unusual orbit suggest a very red dark surface similar to some red Centaurs and TNOs, elongated, most probably non-convex shape, and rather slow rotation.